\documentstyle[11pt]{article} 
\def\ie{{\it i.e.}}
\def\tshalf{ {\textstyle {1\over 2} }}
\def\un{      U($N$) }
\def\uneven{  U($N$ even) }
\def\unodd{  U($N$ odd) }
\def\son{     SO($N$) }
\def\soneven{  SO($N$ even) }
\def\sonodd{  SO($N$ odd) }
\def\soeven{  SO($2n$) }
\def\soodd{   SO($2n+1$) }
\def\spinn{  Spin($N$) }
\def\spineven{  Spin($2n$) }
\def\spinodd{   Spin($2n+1$) }
\def\spn{     Sp($N$)    }
\def\speven{  Sp($2n$)   }
\def\ih{ {(\pm)} }
\def\i{ {(+)} }
\def\h{ {(-)} }
\def\a{\alpha}  
\def\b{\beta}  
\def\d{{\rm d}}
\def\e{{\rm e}}  
\def\Z{{\bf Z}}
\def\be{\begin{equation}} 
\def\ee{\end{equation}} 
\def\bea{\begin{eqnarray}} 
\def\eea{\end{eqnarray}} 
      
\renewcommand{\baselinestretch}{1.0} 
 
\begin{document} 
\begin{flushright} 
hepth/9601104\\
BRX--TH--365\\ 
BOW--PH--106\\ 
BERC--PH--105
\end{flushright} 
 
\begin{center} 
{\Large\bf Evaluation of the Free Energy of \\ 
\vspace{.2in} 
 
Two-Dimensional Yang--Mills Theory\footnote{Research supported in part 
by the DOE under grant DE-FG02-92ER40706} } 
\vspace{.2in} 
 
Michael Crescimanno\footnote{e-mail: crescima@physics.berea.edu}\\ 
Department of Physics\\ 
Berea College\\ 
Berea, KY 40404 
 
\vspace{.15in} 
 
Stephen G. Naculich\footnote{e-mail: naculich@polar.bowdoin.edu}\\ 
Department of Physics\\ 
Bowdoin College\\ 
Brunswick, ME 04011 
 
\vspace{.1in} 
 
and 
 
\vspace{.1in} 
 
Howard J. Schnitzer\footnote{e-mail:  schnitzer@binah.cc.brandeis.edu.}\\
Department of Physics\\ 
Brandeis University\\ 
Waltham, MA 02254 
 
\end{center}

\renewcommand{\baselinestretch}{2} 
\small\normalsize 
 
\begin{center} 
{\bf Abstract} 
\end{center} 
The free energy in the weak-coupling phase 
of two-dimensional Yang-Mills theory on a sphere
for \son~and \spn
is evaluated in the $1/N$ expansion
using the techniques of Gross and Matytsin. 
Many features of Yang-Mills theory are universal
among different gauge groups in the large $N$ limit,
but significant differences arise 
in subleading order in $1/N$.
\vfill 
\eject  
\noindent{\bf I. ~Introduction} 
 
Two-dimensional Yang-Mills theories 
have been used as a laboratory 
to uncover general non-perturbative features  
of gauge theories \cite{Wit}--\cite{GM}.   
It has been shown  
that the $1/N$ expansion of these theories 
may be represented as a formal string theory,  
for SU($N$) and \un~gauge groups \cite{GTM,NR}
as well as for \son~and \spn~\cite{NRS}. 
It is useful to compare these various  string theories
in order to learn 
which  structures are generic,
and what one might expect in a four-dimensional 
string theory of QCD. 
 
Certain features of 2d Yang-Mills theories are
universal, \ie, independent of the gauge group,
in the large $N$ limit \cite{CS}.
For example, the normalized VEV's of Wilson loops
on arbitrary surfaces do not depend 
on the gauge group to leading order in $1/N$,
a fact most naturally understood from the 
string-theoretic interpretation of these theories \cite{CNS}.
On the other hand,
the universality of gauge theory observables 
breaks down in subleading orders of the $1/N$ expansion.
An  example of this is the contribution from cross-caps
which appear on the world-sheet for \son~and \spn, 
but not for SU($N$) or \un~\cite{NRS}. 
It is important to have a clear understanding of the role
of the gauge group in the string interpretation.

To further analyze the differences between these theories,
in this paper we evaluate the 
free energy of Yang-Mills theory on the sphere
in the small area (weak-coupling) phase, 
including exponential corrections to the 1/$N$ expansion.
Our analysis closely parallels that of 
Gross and Matytsin \cite{GM} for U($N$),
focusing specifically on the gauge groups \son~and \spn.
One of the more interesting results of our analysis is the
difference in the double-scaling limit for different gauge
groups (see eq.~(\ref{eq:ds})ff).
\vfil\break

\noindent{\bf II. ~The Partition Function} 

The partition function of two-dimensional 
Yang--Mills theory on the sphere is 
\renewcommand{\theequation}{\arabic{equation}} 
\setcounter{equation}{0} 
\be
Z_0 = \sum_R \; (\dim R)^2 \; \e^{-\frac{\lambda \bar{A}}{2N}C_2(R)} 
\label{eq:partfcn}
\ee 
where the sum 
is over all irreducible representations $R$ of the gauge group, 
$\dim R$ and $C_2(R)$ are the dimension and quadratic Casimir of $R$,
$\bar{A}$ is the area of the sphere,
and $\lambda = e^2 N$, 
where $e$ is the gauge coupling. 
The quadratic Casimir is given by
\be 
C_2(R) = fN \left[ r-U(r) + \frac{T(R)}{N} \right] 
\ee 
with  
\be
f = \left\{ 1 \atop 1/2 \right., \qquad \qquad 
U(r) = \left\{ r/N  \atop -r/N \right.,\qquad  \qquad
\mbox{for} \left\{ \mbox{\son} \atop \mbox{\spn} \right.
\ee
and 
\be
T(R) = \sum^{n}_{i=1} n_i (n_i + 1 - 2i) = 
\sum^{k_1}_{i=1} n^2_i - \sum^{n_1}_{j=1} k^2_j  
\ee 
where $n_i(k_i)$ are the row (column) lengths of the Young diagram 
associated with $R$,  
and $n$ is the rank of the gauge group.
(Our convention is rank \speven $= n$.)
Defining
\bea
\ell_i & = n_i + n -i, \qquad\qquad
   m_i & = n-i 
\qquad \qquad\mbox{for \soeven},  \nonumber\\
\ell_i & = n_i + n - i + \tshalf,\qquad 
   m_i & = n - i + \tshalf 
\qquad \mbox{for \soodd} \\
\ell_i & =  n_i + n - i + 1, \qquad
   m_i & = n - i + 1  
\qquad \mbox{for \speven}    \nonumber
\eea
the dimension and quadratic Casimir of $R$ may be expressed as \cite{FH} 
\be
\dim R   =  \left\{
\begin{array}{ll}
\prod^n_{i<j} \; \frac{(\ell^2_i - \ell^2_j)}{(m^2_i - m^2_j)} 
\qquad & \mbox{ for \soeven }  \\
\prod^n_{i<j} \; \frac{(\ell^2_i - \ell^2_j)}{(m^2_i - m^2_j)} \; 
\prod^n_{i=1} \; \frac{\ell_i}{m_i} 
\qquad & \mbox{ for \soodd and \speven } \\
\end{array}
\right.
\ee
and
\be
C_2(R) = \left\{
\begin{array}{ll}
\sum^n_{i=1} \; \ell^2_i ~ - ~ \frac{1}{24} N (N-1)(N-2)
\qquad & \mbox{for \son} \\
\tshalf \sum^n_{i=1} \; \ell^2_i ~ - ~ \frac{1}{48} N (N+1)(N+2) 
\qquad & \mbox{for \spn}. \\
\end{array}
\right.
\ee
These expressions are also valid for spinor representations of Spin($N$),
which are associated with Young diagrams with $n_i \in {\Z} + \tshalf$.

The partition function eq.~(\ref{eq:partfcn})
depends on the area only through the dimensionless combination 
$ A = \lambda f \bar{A} $
and is given up to an overall constant by
\be
Z_0 (A,N) \propto \left\{
\begin{array}{ll} 
\e^{\b (A,N) }\; X^\i (\a) & \qquad\mbox{for \soeven} \\ 
\e^{\b (A,N) }\; Y^\h (\a) & \qquad\mbox{for \soodd} \\ 
\e^{\b (A,N) }\; Y^\i (\a) & \qquad\mbox{for \speven}\\ 
\e^{\b (A,N) }\; \left[ X^\i (\a) + X^\h (\a) \right]
             & \qquad\mbox{for \spineven} \\ 
\e^{\b (A,N) }\; \left[ Y^\i (\a) + Y^\h (\a) \right]
              & \qquad\mbox{for \spinodd} \\ 
\end{array} 
\right.
\ee
with
\be
\a = {A\over 2N}, \qquad\qquad
\b(A,N)  =  \left\{
\begin{array}{ll} 
{A \over 48} (N-1)(N-2), & \qquad\mbox{for \son and \spinn} \\ 
{A \over 48} (N+1)(N+2), & \qquad\mbox{for \spn}  \\
\end{array} 
\right.  
\ee 
and
\bea
X^\ih (\a) & = &  \sum_{\ell_1 > \ldots\ > \ell_n \geq 0} 
\Delta^2 (\ell^2_1, \ldots , \ell^2_n)\; 
 \e^{-\a\sum^n_{j=1} \, \ell^2_j}  \nonumber\\
Y^\ih (\a) & = &  \sum_{\ell_1 > \ldots\ > \ell_n \geq 0} 
\Delta^2 (\ell^2_1, \ldots , \ell^2_n )  
\left( \prod^n_{i=1} \; \ell^2_i \right)  \e^{-\a \sum^n_{j=1} \ell^2_j} 
\label{eq:XY}
\eea
where the $\ell_i$ are integers in $X^\i$ and $Y^\i$
and half-integers in $X^\h$ and $Y^\h$, 
and
\renewcommand{\baselinestretch}{1} 
\small 
\normalsize 
\be
\Delta (\ell^2_1, \ldots, \ell^2_n ) ~~  = ~~
\prod^n_{i<j} \; (\ell^2_i - \ell^2_j) ~~ = ~~ 
\left| \begin{array}{ccc} 
1 & \ldots & 1 \\ 
\ell^2_1 & \ldots  & \ell^2_n \\ 
\vdots  & \ddots & \vdots \\ 
(\ell^2_1)^{n-1} & \ldots &(\ell^2_n)^{n-1}  
\end{array} \right| 
\ee 
\renewcommand{\baselinestretch}{2} 
\small 
\normalsize 
is the van der Monde determinant in the variables $\ell^2_i$.  
Note that both tensor and spinor representations 
contribute to the partition function for \spinn,
while only tensor representations contribute for \son.
As in the \un case \cite{GM}, 
the expressions eq.~(\ref{eq:XY}) are symmetric 
with respect to the interchange  
$\ell_j \leftrightarrow \ell_k$, 
and vanish when $\ell_j = \ell_k$,
so the summations can be extended to $-\infty < \ell_j < \infty$ 
for all $\ell_j$,  
yielding
\bea
X^\ih (\a) & = &  
{1 \over 2^n ~ n! }
\sum_{-\infty < \ell_1, \ldots, \ell_n < \infty} 
\Delta^2 (\ell^2_1, \ldots, \ell^2_n)\; \e^{-\a \sum^n_{j=1} \ell^2_j} 
\nonumber\\
Y^\ih (\a) & = &  
{1 \over 2^n ~ n! }
\sum_{-\infty < \ell_1, \ldots, \ell_n < \infty} 
\Delta^2 (\ell^2_1, \ldots, \ell^2_n)
\left( \prod^n_{i=1} \; \ell^2_i \right)  
\e^{-\a \sum^n_{j=1} \ell^2_j} 
\label{eq:XYagain}
\eea
where again the $\ell_i$ are integers in $X^\i$ and $Y^\i$
and half-integers in $X^\h$ and $Y^\h$. 

To further evaluate eq.~(\ref{eq:XYagain}),
we introduce several sets of polynomials in $x^2$,
$q^\ih_j (x|\a ) = x^{2j} + \cdots$,
and  
$r^\ih_j (x|\a ) = x^{2j} + \cdots$.
They are defined to be mutually orthogonal 
with respect to the discrete measures
\bea
\sum_x \, \e^{-\a x^2} q^\ih_i (x|\a) q^\ih_j (x|\a)  
&=& \delta_{ij} \, f^\ih_j (\a ) 
\nonumber\\
\sum_x \, \e^{-\a x^2} x^2 r^\ih_i  (x|\a ) r^\ih_j (x|\a ) 
&=&  \delta_{ij} \, g^\ih_j (\a ) 
\label{eq:orthog}
\eea
where the sums on $x$ are over integers 
for $q^\i$ and $r^\i$
and half-integers for $q^\h$ and $r^\h$.
Defining
\bea
q^\ih_j (x|\a ) & =  p^\ih_{2j} (x|\a),  \qquad\qquad
f^\ih_j (\a )  =  h^\ih_{2j} (\a )  \nonumber\\
x r^\ih_j (x|\a ) & =  p^\ih_{2j +1} (x|\a), \qquad\qquad 
g^\ih_j (\a )  =  h^\ih_{2j+1} (\a ) 
\eea
the orthogonality relations eq.~(\ref{eq:orthog}) reduce to
\be
\sum_x \, \e^{-\a x^2} p^\ih_i (x|\a) p^\ih_j (x|\a) = 
\delta_{ij} \, h^\ih_j (\a ) 
\qquad\qquad x \in \left\{ \Z \atop \Z  + \tshalf \right.
\ee 
The $p^\ih_j (x|\a ) = x^j + \cdots$ 
are the polynomials introduced by Gross and Matytsin
\cite{GM} in their study of U($N$).  
They showed that the $h^\ih_j (\a) $ are given by
\be
h^\ih_j(\a ) = h^\ih_0 (\a )\: \prod^j_{i=1} \: R^\ih_i (\a ) ,
\qquad\qquad h^\ih_0(\a) = \sum_{x} \e^{-\a x^2},
\qquad x \in \left\{ \Z \atop \Z  + \tshalf \right.
\label{eq:hR}
\ee
where the $R^\ih_j (\a)$ are defined through the recursion relations
\be
xp^\ih_j (x|\a ) = p^\ih_{j+1} (x|\a ) + R^\ih_j (\a ) p^\ih_{j-1} (x|\a ),
\qquad\qquad R^\ih_0 (\a ) = 0
\ee
and satisfy the differential relations \cite{GM}
\be
{\d \over \d \a} \ln R^\ih_j (\a)
= R^\ih_{j-1} (\a) -  R^\ih_{j+1} (\a),\qquad\qquad
{\d \over \d \a} h^\ih_0 (\a) 
= - R^\ih_1 (\a).
\label{eq:diffeq}
\ee
Rewriting the van der Monde determinants in  eq.~(\ref{eq:XYagain})
in terms of the polynomials $q^\ih_n (x|\a )$ and $r^\ih_n (x|\a)$
and using the orthogonality relations,
we find that the partition function is given up to a constant by
\be
Z_0(A,N) \propto \left\{
\begin{array}{ll}
\e^\b \: \prod^{n-1}_{j=0} \: h^\i_{2j} (\a ) 
\qquad & \mbox{for \soeven} \\
\e^\b \: \prod^{n-1}_{j=0} \: h^\ih_{2j+1} (\a ) 
\qquad &\mbox{for} \left\{ \mbox{\speven} \atop \mbox{\soodd} \right. \\
\e^\b \: \left[ \prod^{n-1}_{j=0} \: h^\i_{2j} (\a ) 
            \quad+\quad \prod^{n-1}_{j=0} \: h^\h_{2j} (\a )  \right]
\qquad & \mbox{for \spineven} \\
\e^\b \: \left[ \prod^{n-1}_{j=0} \: h^\i_{2j+1} (\a ) 
            \quad+\quad  \prod^{n-1}_{j=0} \: h^\h_{2j+1} (\a ) \right]
 \qquad &\mbox{for \spinodd}  \\
\end{array}
\right.
\label{eq:partition}
\ee
The free energy for the orthogonal and symplectic groups is therefore 
\be
F(A,N)  =  \ln \: Z_0  = \b(A,N)  + F_N (A) + \mbox{const}
\ee
with
\be
F_N(A) 
 = \left\{
\begin{array}{ll}
 n \: \ln \: h^\i_0 (\a ) +  \sum^{n-1}_{j=1} \: 
(n-j) \left[ \ln R^\i_{2j-1} (\a ) + \ln \: R^\i_{2j} (\a ) \right] 
 &\mbox{ for \soeven} \\
 n \: \ln \: \left[ h^\ih_0 (\a )   R^\ih_1 (\a )  \right]
 + \sum^{n-1}_{j=1} \: (n-j) 
\left[ \ln R^\ih_{2j} (\a ) + \ln \: R^\ih_{2j+1} (\a ) \right] 
\quad & \mbox{for} \left\{ \mbox{\speven} \atop \mbox{\soodd} \right.  \\
\end{array}
\right.
\label{eq:FSOn}
\ee 
to be compared with the result for U($N$) \cite{GM} 
\be
F_N (A) = N \ln h^\ih_0 (\a) + \sum^{N-1}_{j=1} \: (N-j) \ln R^\ih_j (\a )   
\qquad \mbox{for} \left\{ \mbox{\unodd} \atop \mbox{\uneven} \right. \\
\label{eq:FSUn}
\ee 
Using eq.~(\ref{eq:diffeq}),
we obtain from eqs.~(\ref{eq:FSOn}) and (\ref{eq:FSUn})  
the specific heat capacities 
\be
\frac{\d^2F(A)}{\d A^2} 
 = \left\{
\begin{array}{ll}
\frac{1}{4N^2} \,  [R^\ih_N \; R^\ih_{N-1}]  \qquad   
& \mbox{for} \left\{ \mbox{\soneven} \atop \mbox{\sonodd} \right.  \\  
\frac{1}{4N^2} \,  [R^\i_N \; R^\i_{N+1}]  \qquad  
& \mbox{for \spn} \\ 
\frac{1}{4N^2} \,  [R^\ih_N \;  (R^\ih_{N+1} + R^\ih_{N-1})]  \qquad 
& \mbox{for} \left\{ \mbox{\unodd} \atop \mbox{\uneven} \right. \\
\end{array}
\right.
\label{eq:specheat}
\ee 

To obtain more explicit expressions for the free energies,
one may expand $R_j^\ih (\a)$,
keeping the leading exponential correction, 
\be
R^\ih_j (\alpha)
=   {j \over 2\a} \mp {2\pi^2\over \a^2} \, 
    \e^{-\pi^2/\a}  \, G_j (\a) + \cdots~.
\label{eq:R}
\ee
Gross and Matytsin \cite{GM}
use the recursion relations eq.~(\ref{eq:diffeq}) to show that
\be
G_j(\a) = \oint {\d t \over 2\pi i} 
\left(1 + {1\over t}\right)^n \e^{-2\pi^2 t/\a}
\label{eq:Gcontour}
\ee
with the contour circling $t=0$ and passing to the right of $t=-1$.
This can then be used to evaluate the free energy 
eq.~(\ref{eq:FSUn}) for \un 
below the phase transition\footnote{We correct a 
sign error in ref.~\cite{GM} for even $N$.}
\be
F_N = - {N^2 \over 2} \ln A \pm  2 \, \e^{-{ 2\pi^2N\over A}} \, 
G_N (\a) + \cdots
\qquad \mbox{for} \left\{ \mbox{\unodd}\atop \mbox{\uneven} \right. 
\ee
In the large $N$ limit, the $G_j (\a)$ have the form \cite{GM}
\bea
G_j (\a)
& \approx  &  (-1)^{j+1} \sqrt{ j \over 32\pi n_c^2}
\left( 1- {j\over n_c}\right)^{-1/4} 
\, \e^{- {2\pi^2 N\over A} \left[\gamma\left(j/n_c\right) -1\right]}\nonumber\\
\gamma(x) & = &
\sqrt{1-x} ~-~ {x\over 2} \ln \left( 1+\sqrt{1-x} \over 1-\sqrt{1-x} \right)
\label{eq:G} \\
n_c & = & {\pi^2 \over 2\a}. \qquad 
\nonumber
\eea

Using eqs.~(\ref{eq:R}) and (\ref{eq:Gcontour}),
we calculate the free energy 
for the orthogonal and symplectic groups eq.~(\ref{eq:FSOn})
below the phase transition
\be
F_N  = \left\{
\begin{array}{ll}
\left(- {N^2 \over 4} + {N\over 4} \right) \ln A
+  \e^{-{ 2\pi^2N\over A}} \, [\pm G_{2n}(\a) - I_{2n}(\a)]
+ \cdots 
\qquad & \mbox{for} 
\left\{ \mbox{SO($N=2n$) } \atop \mbox{SO($N=2n+1$)} \right.  \\  
\left(- {N^2 \over 4} - {N\over 4} \right) \ln A
+  \e^{-{ 2\pi^2N\over A}} \, [ G_{2n}(\a) + I_{2n}(\a)]
+ \cdots 
\qquad & \mbox{ for Sp($N=2n$)} \\
\end{array}
\right.
\nonumber 
\ee
where
\be
I_{2n}(\a) 
= -{2 \pi^2  \over \a} \sum_{j=1}^{n} {G_{2j-1} (\a) \over 2j-1}
= \oint {\d t \over 2\pi i} 
\left(1 + {1\over t}\right)^{2n} {\e^{-2\pi^2 t/\a} \over 2t+1}.
\ee
In the large $N$ limit, this yields
\be
F_N  = \left\{
\begin{array}{ll}
\left(- {N^2 \over 4} + {N\over 4} \right) \ln A
 \pm \left( 1- {1 \over \sqrt{1-A/\pi^2} } \right) 
\, \e^{-{2\pi^2N\over A}} \, G_N (\a) + \cdots
\qquad & \mbox{for} \left\{ \mbox{\soneven} \atop \mbox{\sonodd} \right.  \\  
\left(- {N^2 \over 4} - {N\over 4} \right) \ln A
 + \left( 1+ {1 \over \sqrt{1-A/\pi^2} } \right) 
\, \e^{-{2\pi^2N\over A}} \, G_N (\a) + \cdots
\qquad & \mbox{ for \spn} \\
\end{array}
\right.
\ee
but these expressions break down 
if the area $A$ nears the critical area $\pi^2$.
For the \spinn~groups, 
the $O( \e^{-{2\pi^2N\over A}})$ 
correction vanishes 
due to cancellation between the tensor
and spinor representations, 
so that the leading correction is $O( \e^{-{4\pi^2N\over A}})$ 
in that case.

Approaching the phase transition from below 
in the double-scaling limit, defined by
\be
A \to \pi^2 \qquad \mbox{and}\qquad  N \to \infty \qquad \mbox{with} \qquad
N^2 (\pi^2 - A)^3 \equiv g^{-2}_{\rm str} = \mbox{constant}, \;  
\ee 
Gross and Matytsin show that $R_j^\ih(\a)$ behaves as \cite{GM}
\be
R_j^\ih = {n_c^2 \over \pi^2} \mp (-)^j n_c^{5/3} f_1 (x) + O(n_c^{4/3}),
\qquad\qquad x = n_c^{2/3} \, \left( 1 - {j \over n_c} \right),
\qquad\qquad n_c \to \infty
\ee
where $f_1(x)$ obeys the Painlev\'e II equation
\be
f_1^{\prime\prime}  - 4x f_1 - \tshalf \pi^2 f_1^3 = 0.
\ee
Using this, we may show that 
in the double scaling limit
the specific heat capacity eq.~(\ref{eq:specheat}) satisfies
\be
{\d^2F_N\over \d A^2} 
=
{n^4_c \over 4\pi^4 N^2} \; 
\left[ 1 
- \frac{2x}{n^{2/3}_c} 
\, - \, \frac{\pi^4}{2n^{2/3}_c} \; f^2_1(x) 
\pm {\pi^2 \over n_c^{2/3}} f_1^\prime (x)  
+ \ldots \right]_{x = x_N}
\qquad 
\mbox{for} \left\{ \mbox{\son} \atop \mbox{\spn} \right.
\label{eq:ds}
\ee
which has an additional term proportional to $f_1^\prime (x)$ 
compared with \cite{GM}
\be
{\d^2F_N\over \d A^2} 
=
{n^4_c \over 2\pi^4 N^2} \; 
\left[ 1 
- \frac{2x}{n^{2/3}_c} 
\, - \, \frac{\pi^4}{2n^{2/3}_c} \; f^2_1(x) 
+ \ldots \right]_{x = x_N}
\qquad 
\mbox{for \un}
\label{eq:dsun}
\ee
Equation (\ref{eq:ds}) gives the one instanton contribution 
to the specific heat for \son~and \spn~in the double-scaling limit.
The computation of the specific heat for \spinn~is
more complicated due to the contributions
to the partition function
eq.~(\ref{eq:partition})
from both tensor and spinor representations.

\vspace{.2in} 
 
\noindent{\bf III. ~Conclusions} 

Many features of two-dimensional Yang--Mills theory
are universal in the large $N$ limit \cite{CS,CNS},
but differ in subleading order in $1/N$.
In this paper, we have explicitly evaluated the
free energy on the sphere in the weak-coupling phase,
and shown how it compares among the different gauge groups.
The double-scaling limit does not appear to be universal.
Any proposed world-sheet action for 
two-dimensional Yang--Mills string theory 
must accommodate both the universal behavior
as well as the differences among the gauge groups.


\begin{thebibliography}{99} 
\bibitem{Wit}  
E. Witten, Commun. Math. Phys. {\bf 141} (1991) 153; 
M. Blau and G. Thompson, Int. Jour. Mod. Phys. {\bf A7} (1992) 3781. 
\bibitem{GTM} 
D. Gross, Nucl. Phys. {\bf B400} (1993) 161;
J. Minahan, Phys. Rev. {\bf D47} (1993) 3430. 
D. Gross and W. Taylor, Nucl. Phys. {\bf B400} (1993) 81;  
Nucl. Phys. {\bf B400} (1993) 395. 
\bibitem{NR}
S. Naculich and H. Riggs,
Phys. Rev. {\bf  D51} (1995) 4394.
\bibitem{NRS} 
S. Naculich, H. Riggs, and H. Schnitzer, Mod. Phys. Lett. {\bf A8}  
(1993) 2223; Phys. Lett. {\bf B319} (1993) 466; 
Int. Jour. Mod. Phys. {\bf A10} (1995) 2097;
S. Ramgoolam, Nucl. Phys. {\bf B418} (1994) 30. 
\bibitem{HCMR} 
P. Horava, ``Topological Strings and QCD in Two Dimensions," 
NATO Advanced Study Institute Series, vol. 328 (1995),
hep--th/9311156;
``Topological Rigid String Theory and Two-dimensional QCD,''
hep--th/9507060;
S. Cordes, G. Moore, and S. Ramgoolam, ``Large N 2D Yang--Mills Theory and 
Topological String Theory,"  hep--th/9402107. 
\bibitem{RDK} 
B. Rusakov, Phys. Lett. {\bf B303} (1993) 95;
M. Douglas and V. Kazakov, Phys. Lett. {\bf B319} (1993) 219; 
D. Boulatov, Mod. Phys. Lett. {\bf A9} (1994) 365; 
B. Rusakov, Phys. Lett. {\bf B329} (1994) 338;
J. Daul and V. Kazakov, Phys. Lett. {\bf B335} (1994) 371. 
\bibitem{Tay}
W. Taylor, ``Counting Strings and Phase Transitions in 2D QCD," 
MIT--CTP--2297, 
hep--th/9404175; 
M. Crescimanno and W. Taylor, Nucl. Phys. {\bf B437} (1995) 3.
\bibitem{CS} 
M. Crescimanno and H. Schnitzer,  
``Universal Aspects of Two-Dimensional Yang--Mills Theory at Large N,"  
MIT--CTP--2337, BRX--TH--358, hep-th/9501099, 
Int. Jour. Mod. Phys. {\bf A} in press.
\bibitem{CNS} 
M. Crescimanno, S. Naculich, and H. Schnitzer, 
Nucl. Phys. {\bf B446} (1995) 3.
\bibitem{GM} 
D. Gross and A. Matytsin, Nucl. Phys. {\bf B429} (1994) 50. 
\bibitem{FH} 
W. Fulton and J. Harris, {\it Representation Theory: A First Course} 
(New York, Springer-Verlag, 1991). 
\end{thebibliography}
\end{document}